\documentclass[article]{aa}

\usepackage{graphicx,amsmath}
\usepackage{natbib}
\usepackage{bm}
\usepackage{txfonts}
\begin{document}

   \title{Star formation history and metallicity in the Galactic inner bulge revealed by the red giant branch bump}

  \author{F. Nogueras-Lara
          \inst{1}
          \and
           R. Sch\"odel 
          \inst{1}
          \and
          H. Dong
          \inst{1}
         \and
           F. Najarro
          \inst{2}
         \and
          A. T. Gallego-Calvente
          \inst{1}
          \and
          M. Hilker
          \inst{3}
          \and
           E. Gallego-Cano
          \inst{1}
         \and
           S. Nishiyama
          \inst{4}
         \and
          N. Neumayer
          \inst{5}
         \and
          A. Feldmeier-Krause
          \inst{6}
          \and
          J.~H.~V. Girard
          \inst{7}
          \and
          S. Cassisi
          \inst{8}
          \and
           A. Pietrinferni
          \inst{8}          
          }

   \institute{
    Instituto de Astrof\'isica de Andaluc\'ia (CSIC),
     Glorieta de la Astronom\'ia s/n, 18008 Granada, Spain
              \email{fnoguer@iaa.es}
         \and
     Centro de Astrobiolog\'ia (CSIC/INTA), ctra. de Ajalvir km. 4, 28850 Torrej\'on de Ardoz, Madrid, Spain
       \and
      European Southern Observatory (ESO), Karl-Schwarzschild-Straße 2, 85748 Garching, Germany
         \and
      Miyagi University of Education, Aoba-ku, 980-0845 Sendai, Japan
        \and
     Max-Planck Institute for Astronomy, K\"onigstuhl 17, 69117 Heidelberg, Germany
         \and
      The University of Chicago, The Department of Astronomy and Astrophysics, 5640 S. Ellis Ave, Chicago, IL 60637, USA
     \and
 Space Telescope Science Institute, Baltimore, MD 21218, USA
 \and
 INAF - Astronomical Observatory of Abruzzo, Via M. Maggini, sn. 64100 Teramo Italy
      }

   \date{}

 
  \abstract
   {The study of the inner region of the Milky Way bulge is hampered by high interstellar extinction and extreme source crowding. Sensitive high angular resolution near-infrared imaging is needed to study stellar populations and their characteristics in such a dense and complex environment.}
   {We aim at investigating the stellar population in the innermost Galactic bulge, to study the star formation history in this region of the Galaxy.}
   {We used the 0.2$''$ angular resolution $JHK_s$ data from the GALACTICNUCLEUS survey to study the stellar population within two $8.0'\times 3.4'$ fields, about 0.6$^\circ$ and 0.4$^\circ$ to the Galactic north of the Milky Way centre and to compare it with the one in the immediate surroundings of Sagittarius\,A*. We also characterise the absolute extinction and the extinction curve of the two fields.}
   {The average interstellar extinction to the outer and the inner field is $A_{K_s} \sim 1.20 \pm 0.08$\,mag and $\sim 1.48 \pm 0.10$\,mag, respectively.  We present $K_{s}$ luminosity functions that are complete down to at least two magnitudes below the red clump (RC). We detect a feature in the luminosity functions that is fainter than the RC by $0.80\pm0.03$ and $0.79\pm0.02$\,mag, respectively, in the $K_s$ band. It runs parallel to the reddening vector. We identify the feature as the red giant branch bump. Fitting $\alpha$-enhanced BaSTI luminosity functions to our data, we find that a single old stellar population of  $\sim12.8 \pm 0.6 $  \  Gyr and $Z = 0.040 \pm 0.003$ provides the best fit. Our findings thus show that the stellar population in the innermost bulge is old, similar to the one at larger distances from the Galactic plane, and that its metallicity is about twice solar  at distances as short as about 60 pc from the centre of the Milky Way, similar to what is observed at about 500 pc from the Galactic Centre. Comparing the obtained metallicity with previous known values at larger latitudes ($|b|>2^\circ$), our results favour a flattening of the gradient at $|b|<2^\circ$. As a secondary result we obtain that the extinction index in the studied regions agrees within the uncertainties with our previous value of $\alpha = 2.30\pm0.08$ that was derived for the very Galactic centre.}
   
   {}


\titlerunning{Star formation history and metallicity in the galactic inner bulge revealed by the RGBB}
\authorrunning{F. Nogueras-Lara et al.}

   \maketitle
%

\section{Introduction}

Intensive work in the past decade has led to the following approximate picture: About 90\% of the population  of the Milky Way bulge belong to a bar structure. The stellar population is old ($>$10 Gyr), with metallicities ranging from [Fe/H] $\lesssim$ -1.0 to supersolar \citep[e.g.][]{Babusiaux:2010aa,Hill:2011aa,Bensby:2011aa,Ness:2013aa,Johnson:2013aa,Bensby:2017aa}. Enrichment in $\alpha$ elements points towards a rapid formation \citep{Fulbright:2006aa,Fulbright:2007aa}. The relative weight of these populations changes depending on the height above the Galactic plane, and the metal-rich population becomes the dominant population close to the Galactic plane \citep[see review by][and references therein]{Barbuy:2018aa}. However, the study of its central most regions is very complex because of the extremely high interstellar extinction and source crowding.

Some of the most recent and best data on the structure of the Galactic bulge are provided by the VISTA Variables in the Via Lactea (VVV) survey \citep{Minniti:2010fk}. However, in the innermost degree of the Galaxy, the VVV data suffer from seeing-limited angular resolution and strong saturation of point sources, which mean that the completeness limit is close to or, for the innermost fields, even brighter than the RC.

The high angular resolution ($\sim$$0.2''$) GALACTICNUCLEUS survey is a $JHK_s$ survey of the central few thousand square parsecs of the Galactic centre (GC) with the High Acuity Wide-field K-band Imager at the Very Large Telescope (HAWK-I/VLT)  that reaches a few magnitudes deeper than existing seeing-limited surveys \citep{Nogueras-Lara:2018aa}. Thus, it is key to a better understanding of the structure of the innermost parts of our Galaxy. 

In this paper, we analyse and compare three fields of the survey: One of them is centred on the massive black hole, Sagittarius\,A*, and the other two lie in the Galactic bar and/or bulge, at about 0.6$^\circ$ and 0.4$^\circ$ to Galactic north, respectively. We clearly identify a double bump in the luminosity functions of the inner bulge that can be explained as the combination of the RC and the red giant branch bump (RGBB) \citep[see, e.g.][]{Nataf:2011lq,Wegg:2013kx}. 

This feature has not been identified before because of the lack of data with sufficiently high angular resolution and/or wavelength coverage. On the other hand, the point spread function (PSF) photometry that is being carried out on the VVV survey \citep{Alonso-Garcia:2017aa} will improve the situation (not publicly available yet). It is possible to see the detected feature in Fig. 2 of \citet{Alonso-Garcia:2017aa}. This supports the reality of the detection and shows that it is not located only in the fields analysed in this work.
The high dependence of the separation between these two features on metallicity allows us to estimate the metallicity and its gradient in the inner bulge fields under study.

\def\arraystretch{1.2}

\begin{table}
\caption{Summary of HAWK-I observations of the bulge fields F1 and F2.}
\label{obs} 
\begin{tabular}{ccccccc}
 &  &  &  &  &  & \tabularnewline
\hline 
\hline 
 & Date & Filter & Seeing$^a$ & N$^b$ & NDIT$^c$ & DIT$^d$\tabularnewline
 & (d/m/year)  &  & (arcsec) &  &  & (s) \tabularnewline
\hline 
 & 24/07/2015 & $J$ & 0.43 & 49 & 20 & 1.26\tabularnewline
F1 & 20/05/2016 & $H$ & 0.56 & 49 & 20 & 1.26\tabularnewline
 & 20/05/2016 & $K_s$ & 0.54 & 49 & 20 & 1.26\tabularnewline
\hline 
 & 24/07/2015 & $J$ & 0.43 & 49 & 20 & 1.26\tabularnewline
F2 & 26/05/2016 & $H$ & 0.33 & 49 & 20 & 1.26\tabularnewline
 & 14/05/2016 & $K_s$ & 0.58 & 49 & 20 & 1.26\tabularnewline
\hline 
\hline 
 &  &  &  &  &  & \tabularnewline
\end{tabular}
\textbf{Notes.}
$^a$In-band seeing estimated From the PSF FWHM.
measured in long-exposure images. The final angular resolution is $\sim 0.2''$ in all three bands.
$^b$Number of pointings. 
$^c$Number of exposures per pointing. 
$^d$Integration time for each exposure. The total integration time of each observation is given by N$\times$NDIT$\times$DIT. 
 \end{table}

\section{Data}

For this study we used the GALACTICNUCLEUS survey \citep{Nogueras-Lara:2018aa}. The 5 $\sigma$ detection limits of the catalogue are approximately at $J = 22$, $H = 21$ and $K_s = 20$ mag. The photometric uncertainty is below $0.05$ mag at $J = 20$, $H = 17$ and $K_s = 16$ mag, and the zero-point uncertainty is 0.036 mag in all three bands.  

\begin{center}
   \begin{figure}
   \includegraphics[scale=0.55]{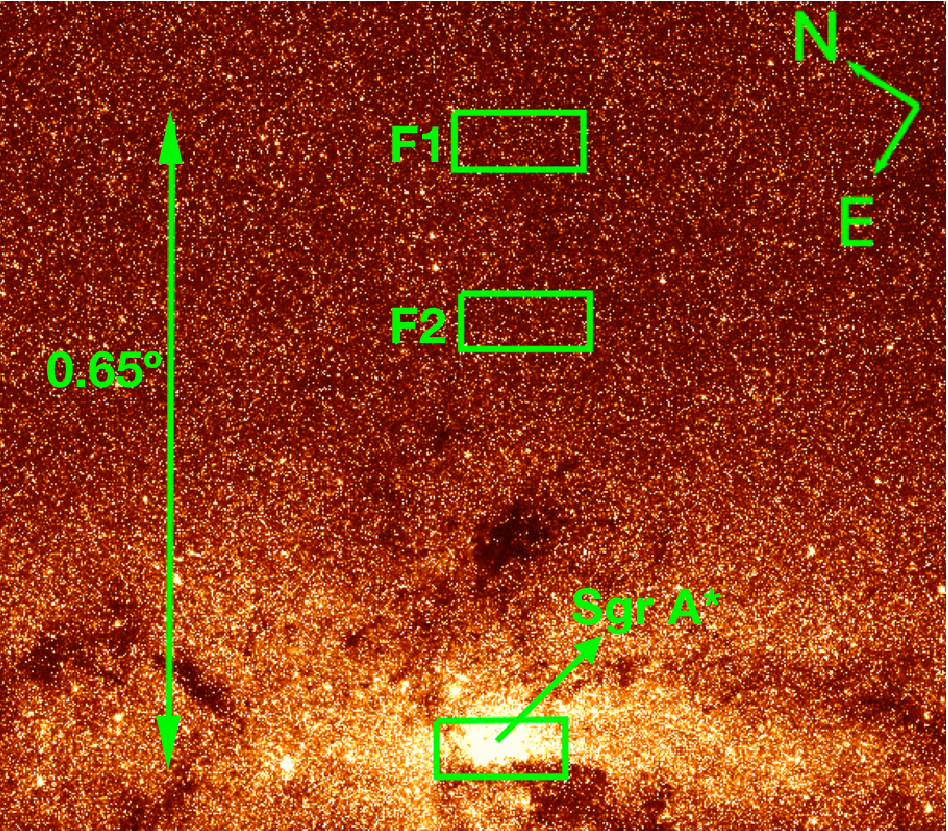}
   \caption{Scheme of the observed fields over-plotted on a Spitzer/IRAC image at 3.6 $\mu$m. F1 and F2 correspond to fields in the bulge, and the field at the bottom is centred on Sagittarius\,A*.}
   
   \label{scheme}
    \end{figure}
    \end{center}

In this paper we use $J$, $H$ and $K_s$ photometry of three fields. A control field (from now on F0), centred on SgrA* (17$^h$ 45$^m$ 40.1$^s$,  -29$^\circ$ 00$'$ 28$''$) and two fields in the bulge (F1 and F2), located $\sim$0.6$^\circ$ and $\sim$0.4$^\circ$ to Galactic North, outside of the Nuclear Bulge (NB) of the Galaxy \citep{Launhardt:2002nx,  Nishiyama:2013uq},  with centre coordinates 17$^h$ 43$^m$ 11.6$^s$,  -28$^\circ$ 41$'$ 54$''$ and 17$^h$ 43$^m$ 53.8$^s$,  -28$^\circ$ 48$'$ 07$''$. The approximate size of the fields is 7.95$'$  $\times$ 3.43$'$. Figure \ref{scheme} shows the location of the fields. Table \ref{obs} summarises the observation information of the fields in the bulge, whereas information on F0 is given in Table\,1 of \citet{Nogueras-Lara:2018aa}.

\section{CMD and identification of a double red clump}

\begin{center}
   \begin{figure}
   \includegraphics[width=\columnwidth]{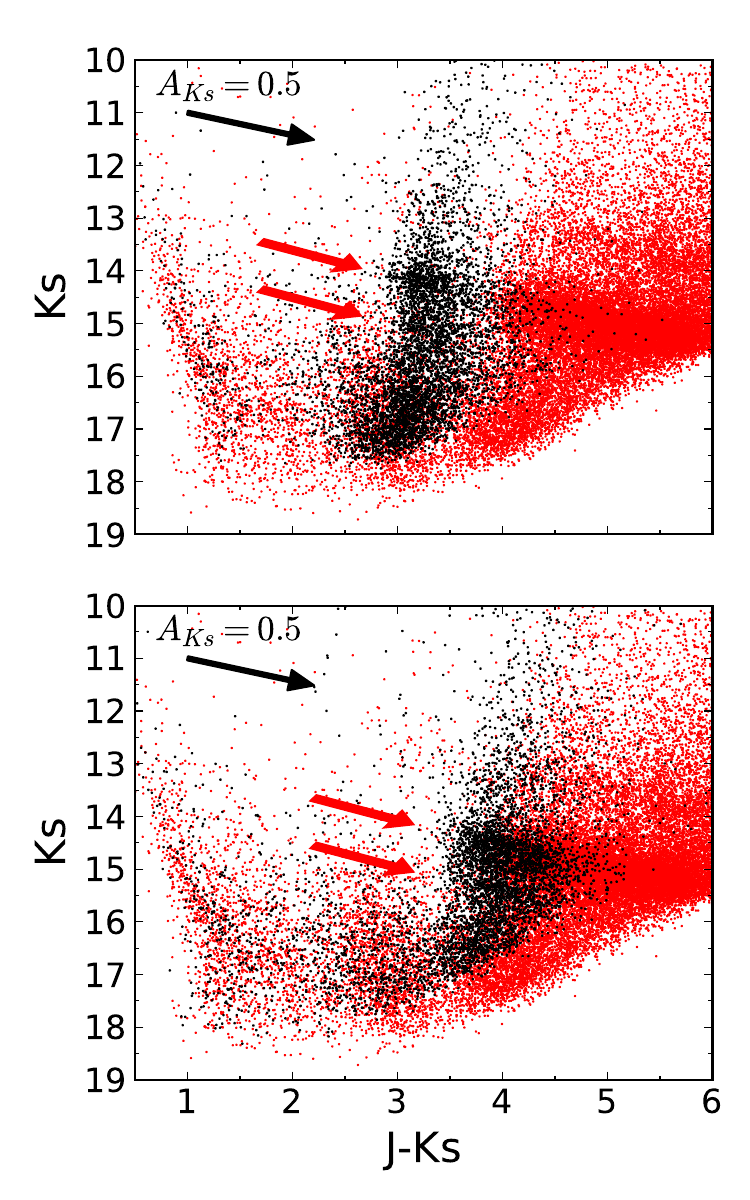}
   \caption{CMD for bulge fields (black) and F0 (red).  The upper and lower panel show F1 and F2, respectively, and the control field, F0. The red arrows indicate the RC in each CMD as well as the fainter density features running parallel to the RC below the red clump. Only a randomly selected fraction of the stars is shown for clarity. The black arrows correspond to an extinction of $A_{Ks} = 0.5$ mag.}
   \label{cmd}
    \end{figure}
    \end{center}

Figure\,\ref{cmd} shows the $J-K_{s}$ colour-magnitude diagram (CMD) of the two fields in the bulge with respect to the control field (in red). Stars with colours $J-K_{s}\lesssim3$ lie in the foreground, with three over-densities indicating spiral arms \citep{Nogueras-Lara:2018aa}. The vertical feature at $J-K_{s}$ between $\approx 3 - 3.5$ in the upper panel corresponds to the bulge stars in F1, while the extended and highly populated feature at $4\lesssim J-K_{s}\lesssim6$ traces the stars in F0. Analogously, the black vertical feature at $J-K_{s}\approx4$ in the lower panel corresponds to the bulge stars in F2. This field has a higher interstellar extinction than F1. The dense regions at $K_{s}\approx14$ and 
$K_{s}\approx15$ indicate the location of RC stars in F1-F2 and F0, respectively. The stars in F0 lie deeply embedded in the central molecular zone, and their extinction is about 1\,mag higher in $K_{s}$ in than in F1. The RC populations of all three fields are aligned following the same reddening vector.

\begin{center}
   \begin{figure}
   \includegraphics[width=\columnwidth]{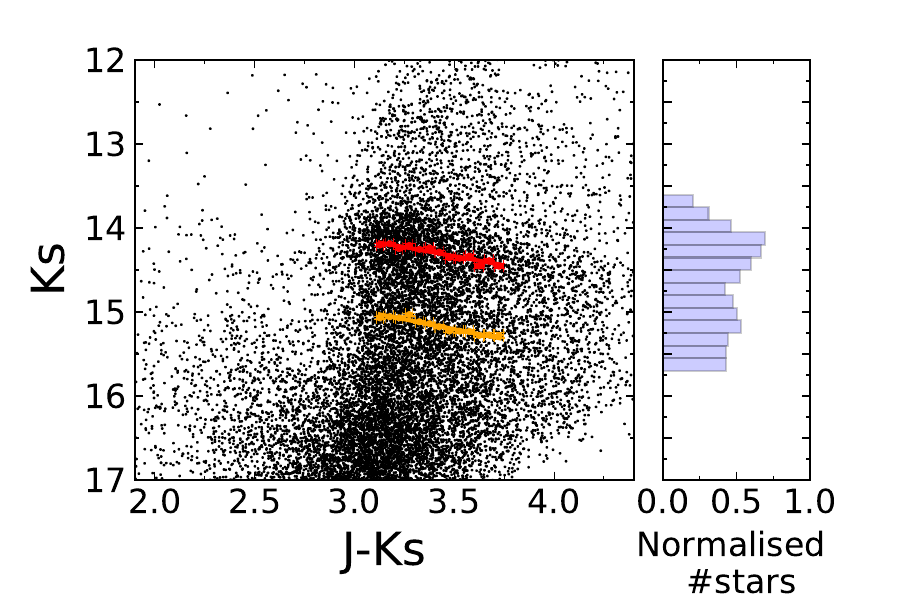}
   \caption{Derivation of the slopes for F1. Main panel: CMD for F1. Red and orange lines indicate the mean values obtained using the GMM and linear fits in several bins 0.05 mag wide in colour. Right panel: Normalised histogram of all the stars in $J-K_s \in [3.1,3.75]$ and $K_s \in [13.6,15.7]$.}
   
   \label{cmd_fit}
    \end{figure}
    \end{center}

A secondary clump is visible at fainter magnitudes below the RC in F1 and F2. To better characterise the visually detected features, we defined a region in the CMD that includes both features ($J-K_s \in [3.1,3.75]$ for F1 and $J-K_s \in [3.6,4.3]$ for F2) and divided it into small bins of 0.05 mag in colour. We analysed the distribution of stars in each bin using the SCIKIT-LEARN python function GaussianMixture \citep[GMM,][]{Pedregosa:2011aa}. In this way, we applied the expectation maximisation algorithm to fit and compared a single- and double-Gaussian model. For this, we used the Bayesian information criterion (BIC) \citep{Schwarz:1978aa} and the Akaike information criterion (AIC) \citep{Akaike:1974aa}. We confirmed the visual detection and obtained that a double-Gaussian model fits the data better. Figure\,\ref{cmd_fit} shows a linear fit to the means of the two Gaussians in each bin. For F1 we obtained a slope of $0.45 \pm 0.05 \pm 0.04$  for the bright and  $0.45 \pm 0.05 \pm 0.02$ for the faint clump. Analogously, we obtained $0.42 \pm 0.03 \pm 0.03$ and $0.43 \pm 0.05 \pm 0.04$ for F2. The first uncertainty corresponds to the statistical uncertainty and was calculated using a jackknife resampling method. The second uncertainty refers to the systematics and was computed considering different bin widths and lower limits of the $J-K_s$ cut-off. Both slopes agree perfectly within their uncertainties.

\section{Characterisation of the features}
 
 \subsection{Interstellar extinction}

In this section we compute the extinction of the two different groups (RC and faint bump) in the two fields and compare it. First of all, for the two fields in the bulge, we computed the stars belonging to each group by obtaining the line that indicates  a\mbox{ 50\%}  probability of membership using the GMM. In addition, to avoid the bias introduced by the initial RC selection, we accepted a star in one of the two groups only if it was within 2 $\sigma$ of the corresponding Gaussian distribution. We built histograms for each feature and obtained a Gaussian-like distribution with mean values of $K_{s\_bright} = 14.28 \pm 0.01 \pm 0.05$ mag and $K_{s\_faint} = 15.15 \pm 0.01 \pm 0.04$ mag for F1 and $K_{s\_bright} = 14.60 \pm 0.01 \pm 0.06$ mag and $K_{s\_faint} = 15.37 \pm 0.01 \pm 0.06$ mag for F2. The errors refer to statistical and systematic uncertainties, respectively. The systematics take into account different cuts of the lower limit of $J-K_s$ and different selections of the bin width. The difference in mean $K_s$ between the two features is thus $0.87 \pm 0.08$ mag and $0.77 \pm 0.08$ mag for F1 and F2, respectively, where the uncertainties were propagated quadratically. If this magnitude difference were only due to extinction, then the fainter clump should have a significantly redder colour. Assuming the extinction curve of \citet{Nogueras-Lara:2018aa}, the corresponding difference in colour would be $\Delta(J-K_{s})\approx1.9\pm0.2$ for both fields. However, the observed colours are similar in both cases.

An alternative way to assess whether interstellar extinction may contribute to the magnitude separation between the features is the following: If both features have approximately the same extinction, then their magnitude separation computed from the difference of the means of the two Gaussians should be equal to the one obtained by averaging over the Gaussian means obtained for each $0.05$\,mag bin that were computed above to compare the slopes of the two features (see Fig. \ref{cmd_fit}). The latter value is $0.86 \pm 0.03$ mag and $0.75 \pm 0.03$ mag for F1 and F2, respectively. The uncertainty refers to the standard deviation of the distribution of the distances. Comparing these two values, we can assume that extinctions between features are very similar in both cases.

Finally, we computed the extinction to each clump from the measured magnitudes of each star. We assumed the extinction index $2.30 \pm 0.08$ derived in \citet{Nogueras-Lara:2018aa}. We computed the theoretical intrinsic colours $J-H$ and $H-K_s$ expected for the stars in the clumps. All these stars lie along a (reddened) red giant branch and will therefore have similar intrinsic colours if their luminosities are not extremely different. Thus, we chose an RC theoretical model to calculate the intrinsic colours following the method described in Sec.\,6.1.\ of  \citet{Nogueras-Lara:2018aa}. We created a grid of extinctions in steps of 0.01 mag and computed the corresponding reddened colours. Then, we defined the $\chi^2$ of the differences between the grid of colours and the real data for each RC feature.  We computed the average extinctions and obtained almost equal values for stars belonging to each feature. The mean values are $A_{K_{s\_bright}} = 1.19 \pm 0.08$ and $A_{K_{s\_faint}} = 1.20 \pm 0.08$ for F1 and $A_{K_{s\_bright}} =1.47 \pm 0.10$ and $A_{K_{s\_faint}} = 1.48 \pm 0.10$, for F2. The uncertainty corresponds to systematics and the statistical uncertainty is negligible. The systematics were computed analogously to those in Sec.\,6.1. of \citet{Nogueras-Lara:2018aa}. 

We conclude that there is no significant difference in absolute interstellar extinction between the two groups of stars. With the obtained absolute extinction values and the slopes of the features, we computed the extinction index using the following expression:

\begin{equation}
\label{eq_slope}
\alpha = -\frac{\log(1+\frac{1}{m})}{\log(\frac{\lambda_{\text{eff}_1}}{\lambda_{\text{eff}_2}})},
\end{equation}

\noindent where $m$ is the slope in the CMD $\lambda_{\rm eff_2}$ versus $\lambda_{\rm eff_1}-\lambda_{\rm eff_2}$, with $\lambda_{\rm eff_1} = J$ and $\lambda_{\rm eff_2} = K_s$ and $\lambda_{\rm eff_i}$ is the effective wavelength \citep[for details, see][]{Nogueras-Lara:2018aa}. For F1, we obtained $\alpha_{bright} = 2.22 \pm 0.15 \pm 0.13$ and  $\alpha_{faint} = 2.21 \pm 0.13 \pm 0.07$, whereas for F2  $\alpha_{bright} = 2.31 \pm 0.09 \pm 0.08$ and  $\alpha_{faint} = 2.28 \pm 0.15 \pm 0.13$. These values agree perfectly with the extinction index of $2.30 \pm 0.08$ obtained in \citet{Nogueras-Lara:2018aa}. Therefore we can conclude that the extinction curve in the near-infrared between $J$ and $K_s$ does not vary spatially between these fields within the uncertainties. The higher uncertainties found for F1 can be explained by the lower extinction of this field. The higher the extinction, the wider the spread of the stars along the reddening vector, which facilitates calculating the slope of the distribution.

\subsection{Extinction map}

To better characterise the detected features, we need to produce extinction maps that allow us to correct the extinction (and the differential extinction) in the studied fields. We calculated extinction maps using the method described in \citet{Nogueras-Lara:2018aa}.

We defined a pixel scale of 0.5''/pixel and used the following equation to compute the extinction:

\begin{equation}
\label{ext_map}
ext = \frac{m_1-m_2-(m_1-m_2)_0}{\left(\frac{\lambda_{m_1}}{\lambda_{m_2}}\right)^{-\alpha}-1} \hspace{0.5cm},
\end{equation}

\noindent where $m_1$ and $m_2$ are the magnitudes for two bands, the subindex 0 indicates the intrinsic colour, and $\lambda_i$ are the effective wavelengths. We used the colour $J-K_s$ since the difference in wavelength is larger than between $H-K_s$ or $J-H$. In this way, we reduced by a factor of three the systematic uncertainty of the map associated to the uncertainty of the ZP in comparison with using $H-K_s$. Moreover, the relative uncertainty of the intrinsic colour is much lower in the case of $J-K_s$ because the uncertainties of the intrinsic magnitudes are of the same order, but the colour term $(J-K_s)$ is $\sim$6 times larger than for $(H-K_{s})$.

To build the extinction map, we used only the stars in the two features. They are most probably RC stars or RGBB stars and have similar intrinsic colours (see Table \ref{basti}). We computed the extinction for each pixel taking into account only the ten closest stars (within a maximum radius of  15'') and weighting the distances with an inverse-distance weight (IDW) method (see \citealt{Nogueras-Lara:2018aa} for details). We did not assign any value to the pixels without the minimum number of required close stars. For the subsequent analysis, we excluded stars located in regions where the extinction maps have no value ($\sim 20 \%$ of the image for both fields). We obtained fairly homogenous extinction maps for F1 and F2 with mean extinctions of $A_{K_s} = 1.14$\,mag and $A_{K_s} = 1.39$\,mag and a standard deviation of $0.06$\,mag and $0.06$\,mag, respectively. The statistical and systematic uncertainties for the extinction maps are $\sim 3 \%$ and  $\sim 7 \%$ for F1 and $\sim 3 \%$ and  $\sim 6 \%$ for F2. The statistical uncertainty considers the dispersion of the values of the ten closest stars and the possible variation of the intrinsic colour, whereas the systematics take into account the uncertainties of the extinction index, the ZP and the effective wavelengths.

\begin{center}
   \begin{figure}
   \includegraphics[width=\columnwidth]{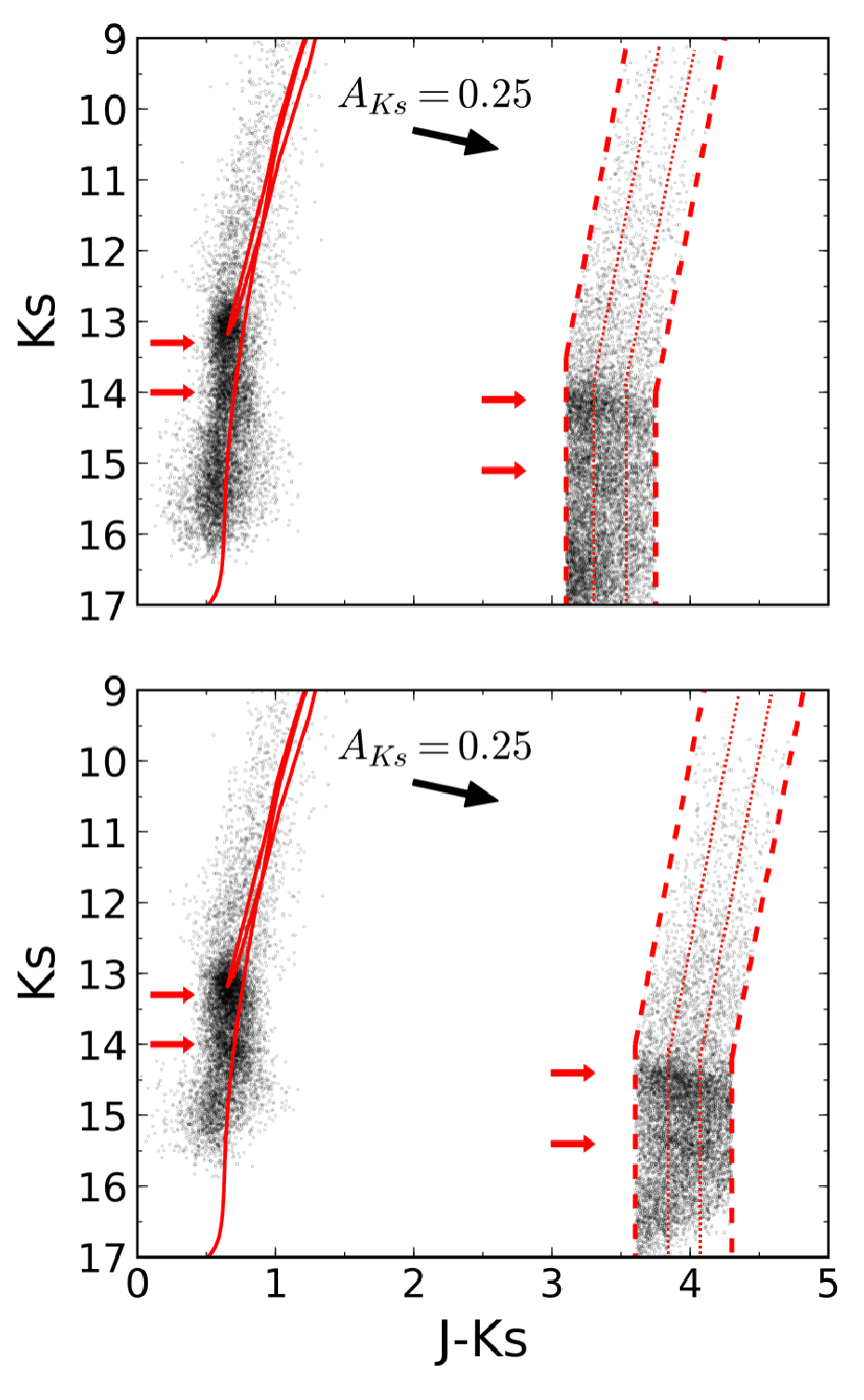}
   \caption{CMD $K_s$ vs. $J-K_s$ with and without extinction correction. Red arrows show the position of the detected features. The black arrow indicates the reddening vector. The red continuous line corresponds to an isochrone of 13.5 Gyr and Z = 0.04 plotted over the de-reddened stars. Upper and lower panels show F1 and F2, respectively. Red dotted lines show different divisions of the data used to produce LFs. Accounting for atomic diffusion would reduce the age of the depicted isochrone $\sim 0.7-1$ Gyr (see main text).}
   
   \label{deredden}
    \end{figure}
    \end{center}

Figure \ref{deredden} shows the extinction-corrected CMDs $K_s$ versus $J-K_s$ for both fields. To exclude the foreground population and highly extinguished or intrinsically reddened stars, we selected only stars between the red dashed lines in the uncorrected CMDs, as indicated in the figure. As can be seen, the application of the extinction map considerably reduces the scatter of the points since it lets us correct the differential extinction. The standard deviation of the distribution of the de-reddened colours around the detected features is $\sigma = 0.07$ in both cases.

\subsection{Luminosity function}

\begin{center}
   \begin{figure}
   \includegraphics[width=\columnwidth]{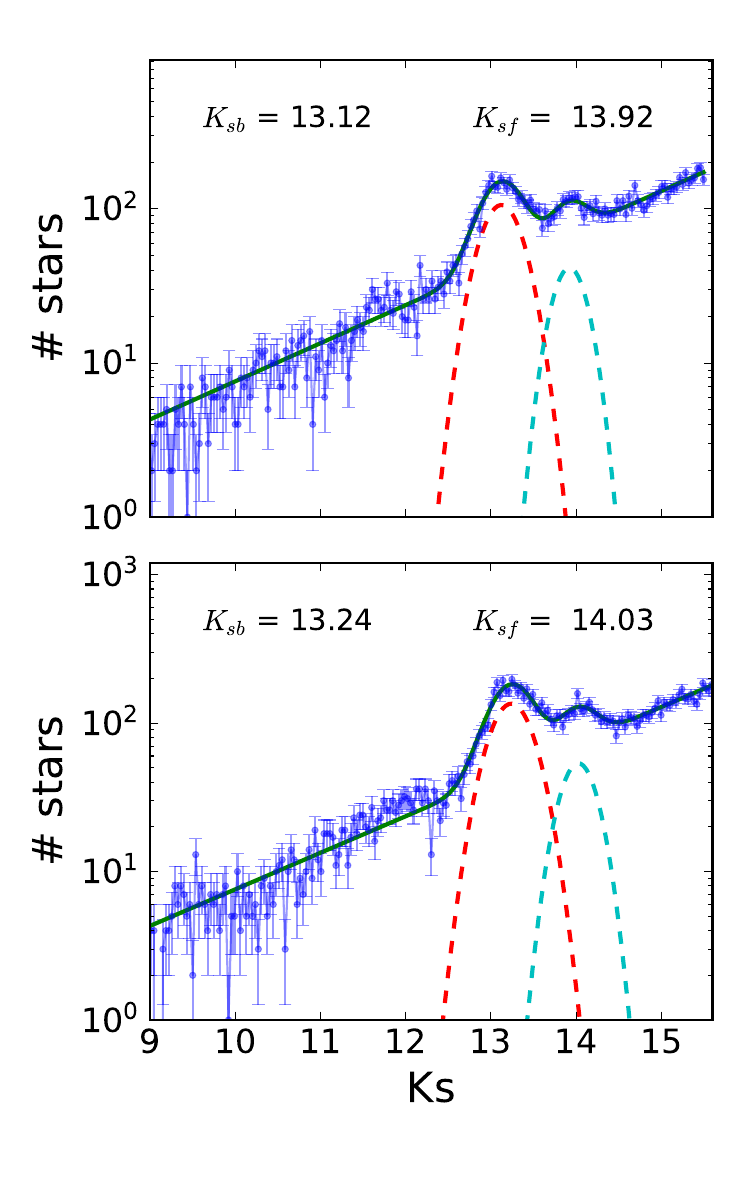}
   \caption{$K_s$-band de-reddened LF. The green line depicts the best fit using an exponential plus two Gaussians model. Red and cyan dashed lines show the position of the two Gaussians of the model. $K_{sb}$ and $K_{sf}$ correspond to the Gaussian peaks of the bright and faint feature, respectively. Upper and lower panel show F1 and F2, respectively.}
   
   \label{lum_exp}
    \end{figure}
    \end{center}

To obtain the luminosity function (LF) of the de-reddened $K_s$ data,  we used the extinction map computed in the previous section and applied it to the $K_s$-band data. To deselect foreground stars, we used the $H-K_{s}$ colour because the $H-$ and $K_{s}-$ band data are far more complete than the $J$-band data due to the steep NIR extinction law. We converted the colour cuts shown in Fig.\,\ref{deredden} into the corresponding colours in $H-K_{s}$ with the extinction curve given by \cite{Nogueras-Lara:2018aa}. Figure\,\ref{lum_exp} shows the obtained LFs, which are complete down to 2\,mag fainter than the faint feature clump.

To determine the reddening-free magnitude of the detected features and the associated distance between them, we fitted a two Gaussian model plus an exponential background to the LFs  \citep[as done in][]{Wegg:2013kx}. We selected a bin width of 0.035 mag to produce the LFs and found that this simple model fits the data well (reduced $\chi^2$ = 1.39 and 1.61 for F1 and F2 respectively). We obtained $K_{s\ bright} = 13.12 \pm 0.01 \pm 0.08$ mag and $K_{s\  faint} = 13.92 \pm 0.03 \pm 0.08$ mag for F1 and $K_{s\  bright} = 13.24 \pm 0.01 \pm 0.07$ mag and $K_{s\  faint} = 14.03 \pm 0.02 \pm 0.07$ mag for F2. The uncertainties correspond to the statistical uncertainty and the systematics, respectively. The systematics were computed considering the systematic error of the extinction map, the values of the fit using different bin widths to create the LF and the ZP uncertainty. The distance between the two features is $\Delta K_s = 0.80\pm0.03$ mag for F1 and $\Delta K_s = 0.79\pm0.02$ for F2. The uncertainty corresponds to the quadratic propagation of the statistical errors, because the systematics affect both peaks equally.

Moreover, we computed the relative fraction of stars between both features, $f_{f/b}$ (number of stars in the faint feature / number of stars in the bright feature) integrating over the Gaussians corresponding to each individual peak. We obtained $f_{f/b} = 0.32 \pm 0.04$ and $f_{f/b} = 0.33 \pm 0.04$ for F1 and F2, respectively. The uncertainty was calculated through Monte Carlo simulations considering the uncertainties of the parameters of the best fit of the LFs.

\section{Discussion}

We discuss the following possible explanations for the two peaks observed in the $K_s$-band LF: 1) RC stars at different distances. 2) A combination of different ages and/or metallicities of the RC stars. 3) The fainter feature could be explained by the red giant branch bump (RGBB). 

\subsection{Possible scenarios}

1) Different distances between the two observed RCs could explain the detected features. Since we have computed the extinction-corrected magnitudes and distance between the two features, we can use the distance modulus to calculate the distances to each one assuming that both correspond to RC stars. For this, we took the values obtained for F1. We assumed an absolute RC magnitude of $M_{K} = -1.54 \pm 0.04$ as in  \citet{Groenewegen:2008wj} and considered that the difference between the $K$ and the $K_s$ magnitudes is  $\approx 0.01$ \citep{Nishiyama:2006ai}. In addition, we used a population correction factor, $\Delta M_{K_s} = -0.07 \pm 0.07$ \citep{Nishiyama:2006ai}. The extinction-corrected magnitude for the bright peak is $K_{s\_bright} = 13.12 \pm 0.08$ (adding quadratically the systematics and statistical uncertainties), and the difference with respect to the faint peak is $0.80 \pm 0.03$ mag. The obtained distance for the bright clump is $8.3 \pm 0.4$ kpc, fully consistent with the well-determined distance of the GC \citep[$8.32\pm0.07|_{stat}\pm0.14|_{sys}$ kpc,][]{Gillessen:2017aa}. Accordingly, the faint RC feature would be located at a distance of $3.7 \pm 0.4$ kpc beyond the bright clump, clearly beyond the GC. However, this scenario is very unlikely, because we previously determined the extinction to each feature and obtained a very similar value for both features. It is highly improbable that there is no extinction between the GC and more than three kpc beyond it. The situation for F2 is analogous. Moreover, similar detections of a second bump in the LF located at $\sim 0.70-0.74$ mag fainter than the RC have been made previously at larger latitudes ($b>2^\circ$) \citep[e.g.,][]{Nataf:2011aa,Nataf:2013fk,Wegg:2013kx}. In those cases, the distance from the Galactic plane means that a spiral arm beyond the GC is highly improbable. For all of this, we can safely exclude this possibility.

2) The luminosity of RC stars depends on their ages and metallicities. However, this variation can account for at most $0.5$\,mag in $K_{s}$ \citep[see Figure\,6 in][]{Girardi:2016fk} if one of the two stellar populations happens to be very young.  Even then, this is still $\sim 0.2$ mag smaller than the observed difference between the detected features. If we assume that all RC stars in our CMD are older than about 2 billion years, then we obtain an even stronger constraint because then the separation between any two populations cannot be larger than $\Delta K_{s}\approx0.2$. Therefore, we cannot explain the observed feature as a consequence of RC stars with different ages or metallicities.

3) The RGBB is a feature in the CMD of old stellar populations corresponding to the evolutionary stage during which the H-burning shell approaches the composition discontinuity left by the deepest penetration of the convective envelope during the $I^{th}$ dredge-up \citep[and references therein]{Cassisi:1997aa,Salaris:2002ab,Nataf:2014aa}. Since the RGBB brightness depends on the maximum depth attained by the convective envelope and on the chemical profile above the advancing H-burning shell, its brightness depends on the stellar metallicity and age (see \citet{Cassisi:2013aa} for a detailed discussion of this issue). To study the magnitude difference between the RGBB and the RC, we used the BaSTI\footnote{http://basti.oa-teramo.inaf.it} isochrones \citep{Pietrinferni:2004aa,Pietrinferni:2006aa} extended along the asymptotic giant branch (some of the isochrones have been computed specifically for this work, using the same BaSTI code as for the freely available ones$^1$). To simulate the stellar population of the bulge, we considered that it can be modelled by a mostly old, $\alpha$-enhanced system \citep{McWilliam:1994aa,Zoccali:2003aa,Lecureur:2007aa,Melendez:2008kx,Alves-Brito:2010aa,McWilliam:2010aa,Johnson:2014aa}. We constructed the LFs corresponding to a range of metallicities and ages (between 5 Gyr and 13 Gyr) using the BaSTI web tools$^1$. We fitted the LFs with a three-Gaussian model plus an exponential background, which takes into account theasymptotic giant branch bump  (AGBB)\footnote{This feature has not been taken into account for the data analysis as it might be too faint to be observed within the uncertainties.} , the RC, and the RGBB. The results are shown in Table \ref{basti}. For the calculations, we assumed a distance modulus of $\mu = 14.60\pm0.05$ \citep{Gillessen:2017aa}. To take into account the uncertainties caused by the different distances of the stars along the line of sight as well as by the extinction correction, we smoothed the theoretical LF data points with Gaussians, the FWHM of which was a free parameter during the fits. For the uncertainties we also considered the (minor) differences between the $K$ and $K_s$ bands.  Table\,2 summarises the resulting RC and RGBB brightnesses and their relative fractions according to the isochrones of different ages and metallicities. Given that the RGBB is expected to be present in LFs of old population and given the good fit by the isochrone models, we can conclude that the detection of the RGBB is the most plausible explanation for the observed greater faintness of the secondary clump compared to the RC.

\begin{table*}
\caption{Properties of the AGBB, RC and RGBB obtained using the alpha-enhanced BaSTI isochrones.}
\label{basti}

\def\arraystretch{1.4}

\small
\begin{tabular}{cccccccccccc}
\label{models}
 &  &  &  &  &  &  &  &  &  &  & \tabularnewline
\hline 
\hline 
Age & Z & Y & {[}Fe/H{]} & {[}M/H{]} & $K_{AGBB}$ & $K_{RC}$ & $K_{RGBB}$ & $(J-K)_{RC}$ & $(J-K)_{RGBB}$ & $f^{RC}_{RGBB}$ & $\Delta_{RGBB-RC}$\tabularnewline
(Gyr) &  &  &  &  &  (mag)&  (mag)& (mag) & (mag) & (mag) &  &(mag) \tabularnewline
\hline 
& 0.0198 & 0.2734 & -0.29 & 0.06 & $11.62$ & $12.92$ & $13.28$ & $0.65$ & $0.69$ & $0.16\pm0.02$ & 0.36$\pm$0.02\tabularnewline
& 0.03 & 0.288 & -0.09 & 0.26 & $11.52$ & $12.91$ & $13.33$ & $0.66$ & $0.69$ & $0.16\pm0.01$ & 0.42$\pm$0.02\tabularnewline
5& 0.035 & 0.295 & -0.02 & 0.329 & $11.54$ & $12.87$ & $13.33$ & $0.69$ & $0.70$ & $0.15\pm0.01$ & 0.46$\pm$0.01\tabularnewline
& 0.04 & 0.303 & 0.05 & 0.4 & $11.52$ & $12.86$ & $13.42$ & $0.66$ & $0.70$ & $0.16\pm0.01$ & 0.56$\pm$0.02\tabularnewline
& 0.045 & 0.310 & 0.105 & 0.454 & $11.5$ & $12.82$ & $13.43$ & $0.70$ & $0.70$ & $0.16\pm0.01$ & 0.61$\pm$0.01\tabularnewline
& 0.05 & 0.316 & 0.16 & 0.51 & $11.47$ & $12.77$ & $13.43$ & $0.74$ & $0.70$ & $0.15\pm0.01$ & 0.66$\pm$0.01\tabularnewline
\hline& 0.0198 & 0.2734 & -0.29 & 0.06 & $11.63$ & $13.02$ & $13.43$ & $0.64$ & $0.69$ & $0.22\pm0.01$ & 0.41$\pm$0.02\tabularnewline
& 0.03 & 0.288 & -0.09 & 0.26 & $11.58$ & $12.97$ & $13.59$ & $0.65$ & $0.73$ & $0.24\pm0.01$ & 0.62$\pm$0.01\tabularnewline
8& 0.035 & 0.295 & -0.02 & 0.329 & $11.55$ & $12.94$ & $13.52$ & $0.68$ & $0.75$ & $0.21\pm0.01$ & 0.58$\pm$0.01\tabularnewline
& 0.04 & 0.303 & 0.05 & 0.4 & $11.53$ & $12.92$ & $13.63$ & $0.70$ & $0.69$ & $0.23\pm0.01$ & 0.71$\pm$0.01\tabularnewline
& 0.045 & 0.310 & 0.105 & 0.454 & $11.49$ & $12.89$ & $13.67$ & $0.68$ & $0.70$ & $0.23\pm0.01$ & 0.78$\pm$0.01\tabularnewline
& 0.05 & 0.316 & 0.16 & 0.51 & $11.5$ & $12.87$ & $13.67$ & $0.71$ & $0.70$ & $0.22\pm0.01$ & 0.80$\pm$0.01\tabularnewline
\hline& 0.0198 & 0.2734 & -0.29 & 0.06 & $11.64$ & $13.02$ & $13.52$ & $0.65$ & $0.65$ & $0.24\pm0.01$ & 0.50$\pm$0.01\tabularnewline
& 0.03 & 0.288 & -0.09 & 0.26 & $11.59$ & $12.97$ & $13.63$ & $0.70$ & $0.69$ & $0.25\pm0.01$ & 0.66$\pm$0.01\tabularnewline
9& 0.035 & 0.295 & -0.02 & 0.329 & $11.58$ & $12.97$ & $13.62$ & $0.69$ & $0.70$ & $0.24\pm0.0$ & 0.65$\pm$0.01\tabularnewline
& 0.04 & 0.303 & 0.05 & 0.4 & $11.55$ & $12.92$ & $13.68$ & $0.70$ & $0.69$ & $0.24\pm0.01$ & 0.76$\pm$0.01\tabularnewline
& 0.045 & 0.310 & 0.105 & 0.454 & $11.52$ & $12.87$ & $13.72$ & $0.75$ & $0.70$ & $0.25\pm0.01$ & 0.85$\pm$0.01\tabularnewline
& 0.05 & 0.316 & 0.16 & 0.51 & $11.5$ & $12.87$ & $13.72$ & $0.70$ & $0.70$ & $0.24\pm0.01$ & 0.85$\pm$0.01\tabularnewline
\hline& 0.0198 & 0.2734 & -0.29 & 0.06 & $11.66$ & $13.07$ & $13.57$ & $0.65$ & $0.65$ & $0.25\pm0.01$ & 0.50$\pm$0.01\tabularnewline
& 0.03 & 0.288 & -0.09 & 0.26 & $11.6$ & $13.02$ & $13.68$ & $0.65$ & $0.66$ & $0.26\pm0.01$ & 0.66$\pm$0.01\tabularnewline
10& 0.035 & 0.295 & -0.02 & 0.329 & $11.57$ & $12.97$ & $13.67$ & $0.70$ & $0.75$ & $0.26\pm0.01$ & 0.70$\pm$0.01\tabularnewline
& 0.04 & 0.303 & 0.05 & 0.4 & $11.54$ & $12.92$ & $13.72$ & $0.75$ & $0.70$ & $0.26\pm0.01$ & 0.80$\pm$0.01\tabularnewline
& 0.045 & 0.310 & 0.105 & 0.454 & $11.52$ & $12.91$ & $13.72$ & $0.71$ & $0.75$ & $0.25\pm0.01$ & 0.81$\pm$0.01\tabularnewline
& 0.05 & 0.316 & 0.16 & 0.51 & $11.5$ & $12.87$ & $13.77$ & $0.75$ & $0.70$ & $0.24\pm0.01$ & 0.90$\pm$0.01\tabularnewline
\hline& 0.0198 & 0.2734 & -0.29 & 0.06 & $11.67$ & $13.07$ & $13.56$ & $0.66$ & $0.71$ & $0.26\pm0.01$ & 0.49$\pm$0.01\tabularnewline
& 0.03 & 0.288 & -0.09 & 0.26 & $11.6$ & $13.02$ & $13.67$ & $0.70$ & $0.70$ & $0.26\pm0.01$ & 0.65$\pm$0.01\tabularnewline
11& 0.035 & 0.295 & -0.02 & 0.329 & $11.58$ & $12.97$ & $13.73$ & $0.70$ & $0.69$ & $0.27\pm0.01$ & 0.76$\pm$0.01\tabularnewline
& 0.04 & 0.303 & 0.05 & 0.4 & $11.56$ & $12.97$ & $13.77$ & $0.70$ & $0.70$ & $0.26\pm0.01$ & 0.80$\pm$0.01\tabularnewline
& 0.045 & 0.310 & 0.105 & 0.454 & $11.54$ & $12.97$ & $13.82$ & $0.70$ & $0.65$ & $0.27\pm0.01$ & 0.85$\pm$0.01\tabularnewline
& 0.05 & 0.316 & 0.16 & 0.51 & $11.5$ & $12.92$ & $13.77$ & $0.70$ & $0.75$ & $0.25\pm0.01$ & 0.85$\pm$0.01\tabularnewline
\hline& 0.0198 & 0.2734 & -0.29 & 0.06 & $11.7$ & $13.13$ & $13.56$ & $0.64$ & $0.76$ & $0.25\pm0.02$ & 0.43$\pm$0.02\tabularnewline
& 0.03 & 0.288 & -0.09 & 0.26 & $11.61$ & $13.07$ & $13.68$ & $0.65$ & $0.74$ & $0.27\pm0.01$ & 0.61$\pm$0.01\tabularnewline
12& 0.035 & 0.295 & -0.02 & 0.329 & $11.59$ & $13.04$ & $13.75$ & $0.68$ & $0.72$ & $0.27\pm0.01$ & 0.71$\pm$0.01\tabularnewline
& 0.04 & 0.303 & 0.05 & 0.4 & $11.56$ & $12.97$ & $13.77$ & $0.70$ & $0.75$ & $0.27\pm0.01$ & 0.80$\pm$0.01\tabularnewline
& 0.045 & 0.310 & 0.105 & 0.454 & $11.54$ & $12.97$ & $13.82$ & $0.70$ & $0.70$ & $0.27\pm0.01$ & 0.85$\pm$0.01\tabularnewline
& 0.05 & 0.316 & 0.16 & 0.51 & $11.5$ & $12.92$ & $13.82$ & $0.75$ & $0.70$ & $0.26\pm0.01$ & 0.90$\pm$0.01\tabularnewline
\hline& 0.0198 & 0.2734 & -0.29 & 0.06 & $11.73$ & $13.24$ & $13.66$ & $0.58$ & $0.71$ & $0.26\pm0.02$ & 0.42$\pm$0.02\tabularnewline
& 0.03 & 0.288 & -0.09 & 0.26 & $11.64$ & $13.12$ & $13.72$ & $0.65$ & $0.70$ & $0.27\pm0.01$ & 0.60$\pm$0.01\tabularnewline
13& 0.035 & 0.295 & -0.02 & 0.329 & $11.6$ & $13.07$ & $13.77$ & $0.65$ & $0.70$ & $0.28\pm0.01$ & 0.70$\pm$0.01\tabularnewline
& 0.04 & 0.303 & 0.05 & 0.4 & $11.58$ & $13.02$ & $13.82$ & $0.71$ & $0.70$ & $0.28\pm0.01$ & 0.80$\pm$0.01\tabularnewline
& 0.045 & 0.310 & 0.105 & 0.454 & $11.55$ & $13.02$ & $13.79$ & $0.70$ & $0.78$ & $0.27\pm0.01$ & 0.77$\pm$0.02\tabularnewline
& 0.05 & 0.316 & 0.16 & 0.51 & $11.53$ & $12.97$ & $13.82$ & $0.70$ & $0.75$ & $0.26\pm0.01$ & 0.85$\pm$0.01\tabularnewline
\hline\end{tabular}

\vspace{0.75cm}

\textbf{Notes.}
$K_{AGBB}$, $K_{RC}$ , and $K_{RGBB}$ are the $K$-band peaks obtained for the Gaussian fits of the LF. $f^{RC}_{RGBB}$ is the relative fraction of RGBB stars of that in RC stars. $\Delta_{RGBB-RC}$ is the distance between the peaks $K_{RGBB}-K_{RC}$. The uncertainties that are not specified in the table are $\Delta K_{AGBB}=\pm 0.05$, $\Delta K_{RC}=\pm0.05$,  $\Delta K_{RGBB}=\pm 0.05$, $\Delta (J-K)_{RC}=0.01,$ and $\Delta (J-K)_{RGBB}=0.01$. Accounting for atomic diffusion would reduce the age of the models $\sim 0.7-1$ Gyr (see main text).
 \end{table*}

$\Delta_{RGBB-RC}$ is strongly  dependent on the metallicity \citep{Nataf:2014aa}, and $f^{RC}_{RGBB}$ is a good indicator of age, as can be seen from the Table \ref{basti}. We can thus use the values measured by us,  $f^{RC}_{RGBB}=0.33\pm0.04$ and $\Delta_{RGBB-RC}=0.80\pm0.02$,
to  constrain the age and metallicity of the inner bulge. We can exclude with more than 4 $\sigma$ significance all scenarios with metallicities below $Z = 0.035$. Similarly, ages younger than about 9\,Gyr can be excluded at a $\gtrsim3$ $\sigma$ level. The best fits are obtained for ages $\gtrsim10$\,Gyr and metallicity $Z=0.04,$ which corresponds to twice solar metallicity.

\begin{center}
   \begin{figure}
   \includegraphics[width=\columnwidth]{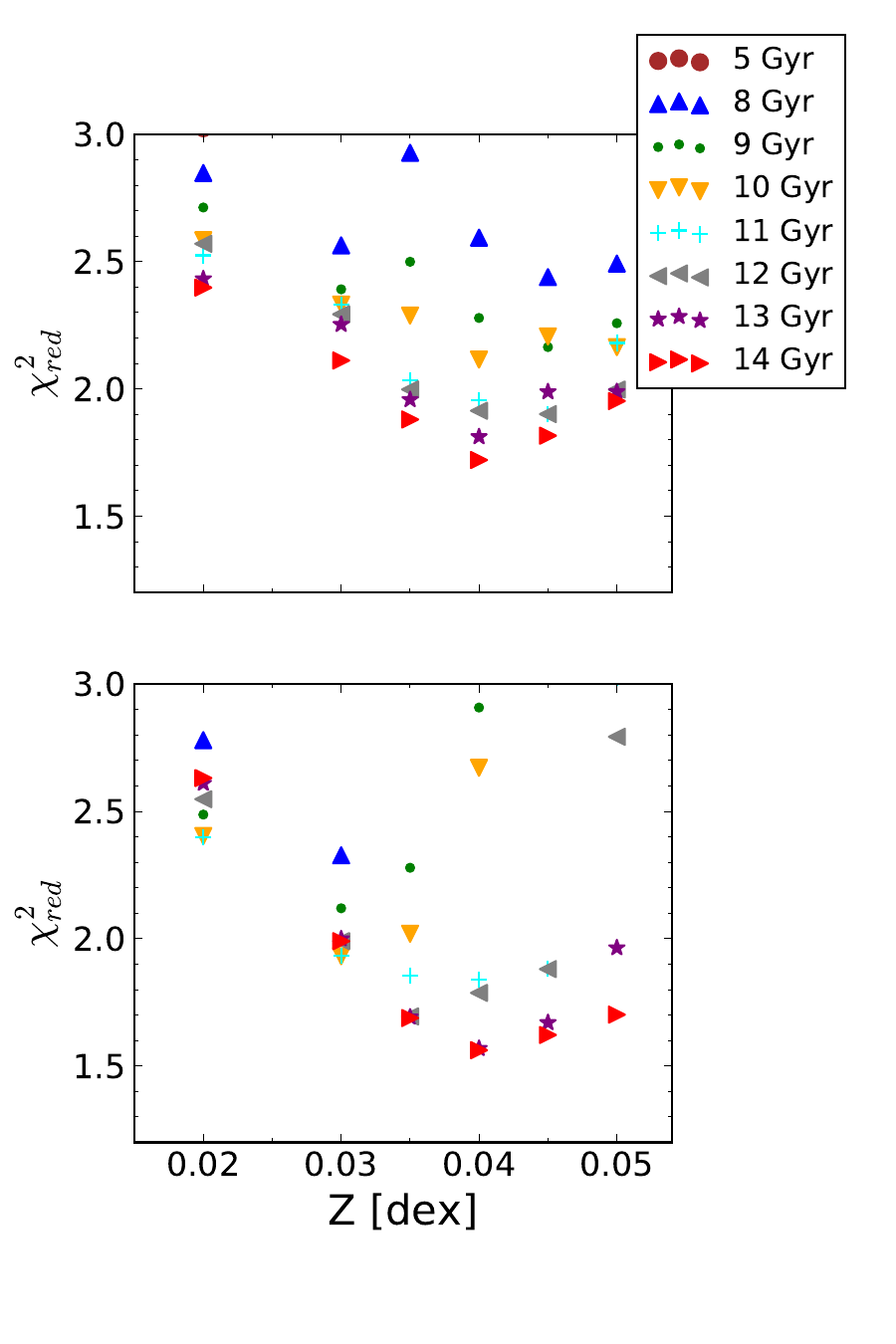}
   \caption{Distribution of the reduced $\chi^2$ for F1 and F2 (upper and lower panel, respectively). No points are associated with 5 Gyr since they have a large reduced $\chi^2$ and the scale is optimised for lower values of the reduced $\chi^2$. Accounting for atomic diffusion would reduce the age of the models $\sim 0.7-1$ Gyr (see main text).}
   
   \label{chi2}
    \end{figure}
    \end{center}

\begin{center}
   \begin{figure}
   \includegraphics[width=\columnwidth]{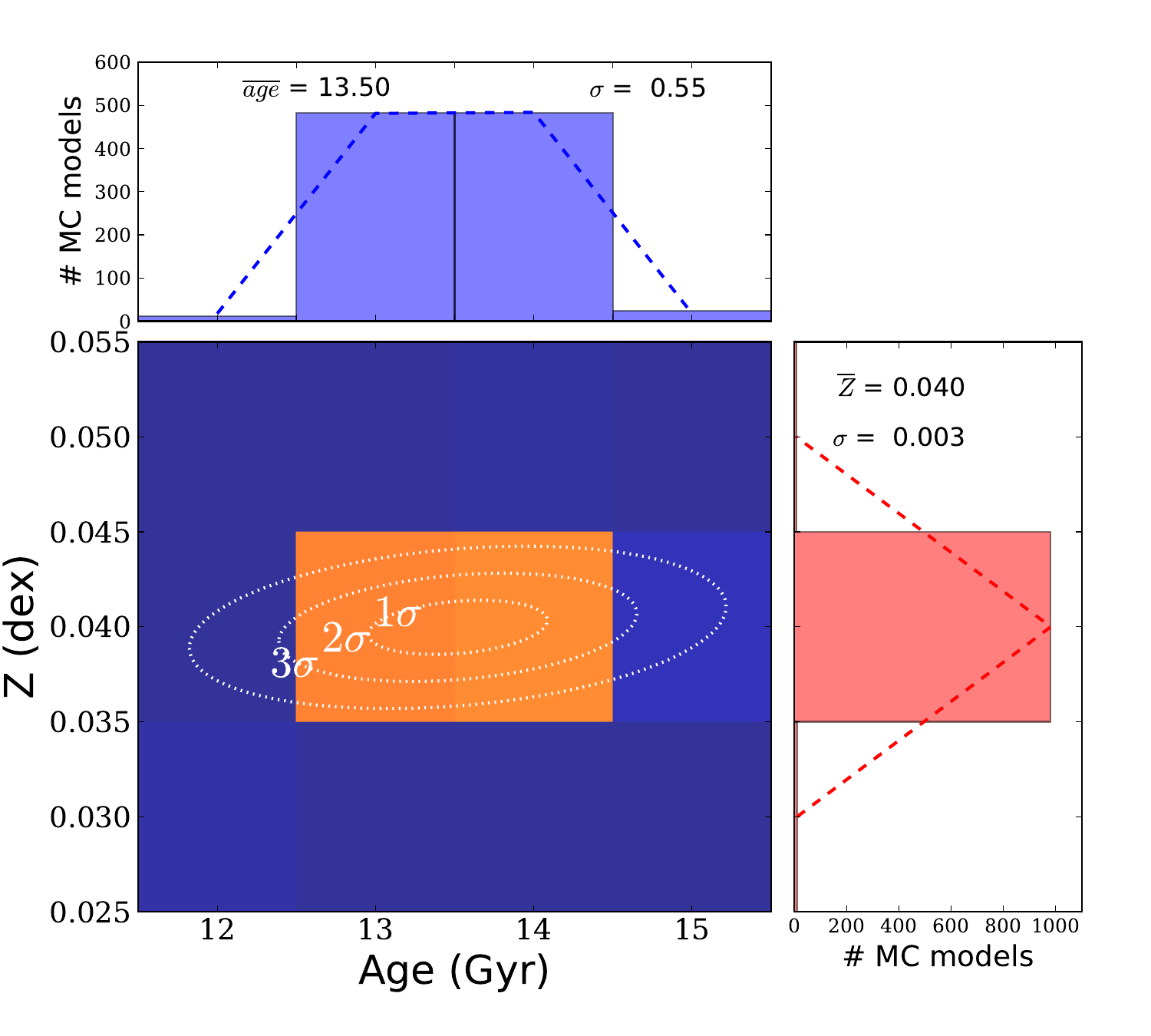}
   \caption{Distribution of ages and metallicities for F2 obtained by fitting 1000 MC samples with the theoretical BaSTI LFs. Upper panel: Age distribution. Right panel: Metallicity distribution. Central panel: Density map of ages and metallicities. Sigma contours are over-plotted in white. Dashed lines show Gaussian fits for ages and metallicities with the values specified in the panels.  Accounting for atomic diffusion would reduce the age of the models $\sim 0.7-1$ Gyr (see main text).}
   
   \label{F2_dist}
    \end{figure}
    \end{center}

\begin{center}
   \begin{figure}
   \includegraphics[width=\columnwidth]{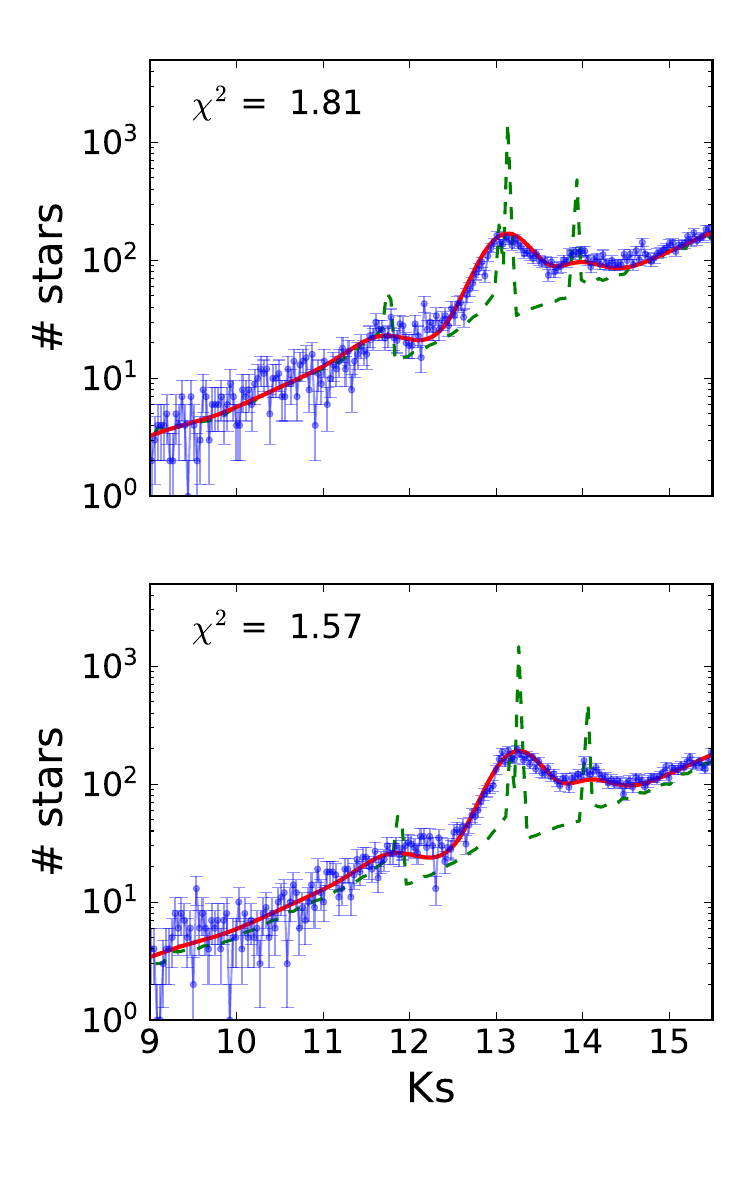}
   \caption{$K_s$-band de-reddened luminosity function. The upper and lower panel show F1 and F2, respectively. The red line depicts the best fit. It is a smoothed 13.5 Gyr, Z = 0.04 LF.  Considering the atomic diffusion in the model would reduce the age $\sim 0.7-1$ Gyr (see main text). The green dashed line corresponds to the LF from BaSTI models, without applying any smoothing.}
   
   \label{LF_basti}
    \end{figure}
    \end{center}

We also fitted  the $K_s$ LFs directly with the  theoretical LFs obtained from BaSTI minimising $\chi^2 = \sum (data-model)/\sigma^2$. The width of the Gaussian to smoothen the theoretical LFs and the distance modulus were set as free parameters. The distance modulus was constrained to lie within 5 $\sigma$ of the expected value $\mu = 14.60\pm0.05$ to avoid false minima in the model fits caused by unphysical values of $\mu$. Figure \ref{chi2} shows the distribution of reduced $\chi^2$ obtained by comparing models with different ages and metallicities with the observed data. A clear minimum appears for F1 and F2 at Z = 0.04 and ages $\sim 13-14$ Gyr. To estimate the uncertainties, we used a Monte Carlo (MC) simulation generating 1000 synthetic LFs from the real data and errors. We fitted the LFs in the same way as the real data using a range of ages from 12 to 15 Gyr (in steps of 1 Gyr) and metallicities Z = 0.03, 0.04 and 0.05. The models we employed do not account for atomic diffusion \citep{Pietrinferni:2004aa,Pietrinferni:2006aa}.  This effect is important for isochrones with ages greater
than a few Gyr \citep{Pietrinferni:2004aa}.  Including atomic diffusion
reduces the age of the models $\sim 0.7 - 1$ Gyr \citep[see sec. 3.9.6,][]{Cassisi:2013aa}, and applying this correction makes our analysis compatible with the
age of the Universe as derived from modelling of the cosmic microwave
background \citep[e.g.,][]{Bennett:2013aa}.   Although Figure \ref{F2_dist} for instance illustrates
an unphysically old age for the models, the correction should be
considered to be implied in that figure, and in the associated analysis. Figure \ref{F2_dist} shows the results of the MC analysis for F2; the result for F1 is similar. Moreover, we estimated the systematic uncertainty introduced by the bin width selection by repeating the fit using a range of different bin widths. Finally, we obtained $13.64\pm0.35$ Gyr and Z = $0.040\pm0.003$ and $13.50\pm0.74$ Gyr and Z = $0.040\pm0.003$ for F1 and F2, respectively. Taking into account the atomic diffusion, we estimate a final age of $12.8 \pm 0.4$  and $12.7 \pm 0.8$ for F1 and F2, respectively (1 $\sigma$ uncertainty). The final uncertainty was obtained considering that the atomic diffusion contributes reducing the ages by $0.85\pm0.15$, and propagating the uncertainties quadratically. Figure \ref{LF_basti} shows the best fits. We conclude that the observed LFs can be satisfactorily modelled by an old single-age population with Z = $0.0400 \pm 0.003$  (or $[Fe/H]= 0.05\pm 0.04$ dex). The statistical and systematic uncertainties have been propagated quadratically. It is important to note that the theoretical LFs used probably also contain systematic uncertainties whose magnitude is difficult to estimate. To check the obtained results, we used the CMD\,3.0 tool (http://stev.oapd.inaf.it/cgi-bin/cmd) \citep{Bressan:2012xy,Chen:2014uq,Tang:2014rm,Marigo:2008aa,Girardi:2010aa} to fit PARSEC evolutionary tracks (version 1.2S) to our data. The best fit was found for an old stellar population model (12 Gyr, which was the oldest population used) with twice solar metallicity. This is in perfect agreement with the result obtained using the BaSTI models.

Accordingly,  the density of the nuclear bulge stellar population in F1 and F2 amounts to approximately 15\% and 30\%, respectively, of the central density of the NB. As a result of factors such as a significantly different  foreground extinction towards the NB and bulge fields and a more complex stellar population in the NB, correcting for any potential bias would be a  complex procedure prone to systematic uncertainties. Nevertheless, while this caveat should be kept in mind, we believe that it does not significantly affect our results because the potential contamination is fairly low for field F1 and because the result of our analysis of field F2 is fully consistent with the one for F1.

The fields we studied here have hardly been investigated before. \citet{Figer:2004fk} studied several fields in the nuclear stellar disc with NICMOS/HST observations. They found that the LFs of most fields provide evidence for continuous star formation in the Galactic centre. Only one of their fields, denominated zc, lies at 0.3 degree to the Galactic north of the nuclear stellar disc. This field alone can be compared to the fields studied here. \citet{Figer:2004fk} found fewer bright stars in this field than in those closer to the GC, in agreement with an older population. The main sequence turn-off in field zc also broadly agrees with a stellar population older than in their other fields. We note that this is only evident if the significantly different extinction towards the different fields is taken into account. \citet{Pfuhl:2011uq} analysed spectroscopic observations of a few hundred giants within 1 pc of Sagittarius A*. They adopted metallicity measurements from other authors and found that at least 80\% of the stellar mass formed more than 5 Gyr ago. However, an important caveat when comparing the results of \citet{Pfuhl:2011uq} to ours is that the stars analysed in the former are all located within the nuclear star cluster of the Milky Way. This has a complex star formation history and even shows evidence of very recent star formation. Stellar populations in nuclear clusters should probably not be compared with those in the surrounding bulges \citep[see, e.g.][]{Neumayer:2017aa,Boker:2010ys}.

\subsection{Fraction of young stars}

Previous studies in the bulge have found a significant fraction of stars with ages $< 5$ Gyr \citep[e.g.][]{Bensby:2013aa,Bensby:2017aa}. Although a single stellar population gives a satisfactory fit to our data, we also repeated the analysis assuming two stellar populations of twice solar metallicity: a young (0.5, 1, 2, 3, 4 and 5 Gyr) and an old population (from 7 to 14 Gyr in steps of 1 Gyr). We found that the young component does not contribute significantly. The $\chi^2$ values do not improve significantly over those obtained with single-population fits. This is in agreement with \cite{Clarkson:2011ys}, who reported that the young stellar population (studied at $b=1.25^\circ$) can be at most $\sim3\%$ and it is also compatible with zero. The more recent work by \cite{Renzini:2018aa} also constrained the contribution of young, high-metallicity stars to $\lesssim3\%$.

\subsection{Spatial variability}
   
We also analysed the age and metallicity variability with the position in the two studied fields. For each field, we randomly selected 30 regions of $\sim 1.8 '$ of radius, where we found enough stars to produce a complete LF. Then, we analysed the obtained LFs following the procedure described in the previous section. The ages of all the random regions follow a quasi-Gaussian distribution with a mean of 13.94 and 13.65 Gyr, and a standard deviation of 0.15 and 0.35 Gyr for F1 and F2 respectively. We did not observe any variation in metallicity and obtained a constant value of twice solar metallicity, in agreement with the results obtained in the previous section.

\subsection{Variation with extinction} 

To study the posible influence of the extinction and different distances to the stars analysed (depth of the bulge), we divided the stars in the CMDs of the two fields into three different sub-sets as shown by the red dashed lines in Fig. \ref{deredden}. We de-reddened each sub-set using the derived extinction map and built LFs. We again fitted all the LFs with the theoretical models and found that there is no significant variation within the uncertainties for ages and metallicities in either field. For F1, the best fit was a star population of 14 Gyr and Z = 0.04 in all three cases, whereas for F2, we found two cases with 14 Gyr and one with 13 Gyr; the metallicity was always Z = 0.04.

 \subsection{Metallicity gradient}

An RGBB feature $\sim 0.70-0.74$ mag fainter than the RC was identified in fields located at vertical distances of $b>2^\circ$ from the GC \citep[e.g.][]{Nataf:2011aa,Nataf:2013fk,Wegg:2013kx}. According to Table\,\ref{models}, this separation implies a metallicity of $[Fe/H]\sim 0$ dex, which is in agreement with the metallicity maps shown for these latitudes by \citet{Gonzalez:2013aa}. Based on these maps, we assumed a metallicity of $[Fe/H]\sim 0$ dex at a vertical distance of 2.5$^\circ$ from the GC. This means a distance of $\sim 300$ pc with respect to F2. Then, we computed the expected metallicity for F2 assuming measured values of the vertical metallicity gradient. We used 0.28, 0.45, and 0.6 dex/kpc \citep{Zoccali:2008aa,Ness:2013aa,Gonzalez:2013aa}. We obtained an expected metallicity of $[Fe/H]$  $\sim 0.08$, $\sim 0.14$ and $\sim 0.18$ dex, respectively. Thus, our result favours a lower metallicity gradient for the regions in the inner bulge. This is in agreement with the flattening of the metallicity gradient in the inner regions inferred by \citet{Rich:2007aa}. This is also compatible with the more prominent fraction of metal-rich stars found close to the plane in comparison with the metal-poor ones and a significantly larger scale height of the latter population, so that its contribution does not vary significantly at small Galactic latitudes \citep{Ness:2013aa,Rojas-Arriagada:2014rt,Barbuy:2018aa}.

\begin{center}
   \begin{figure}
   \includegraphics[width=\columnwidth]{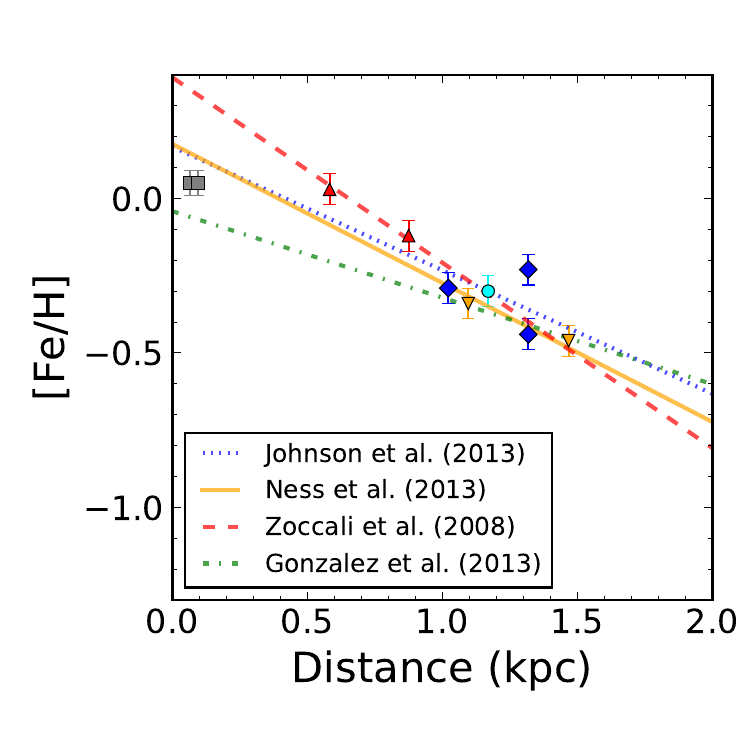}
   \caption{Metallicity measurements at different latitudes reported by previous studies: red triangles correspond to (l, b) = (+1, -4) and (0, -6) fields from \citet{Zoccali:2008aa}, blue diamonds are from \citet{Johnson:2013aa} and correspond to (l, b) = (8.5, 9), (-5.5, -7) and (-4, -9), the inverted yellow triangles correspond to (l, b) = (0, -7.5) and (0, -10) from \citet{Ness:2013aa}, the cyan circle corresponds to  (0, -8) from \citet{Johnson:2011aa},  and the grey squares are from this work. The lines show several metallicity gradients obtained by the authors in the legend. The lines and points share the same colour if they are part of the same work.}
   
   \label{met_gradient}
    \end{figure}
    \end{center}

Figure\,\ref{met_gradient} shows the median values of the metallicities obtained by several authors at different latitudes. We assumed symmetry with respect to the Galactic plane, in agreement with Fig.\,7 of \citet{Gonzalez:2013aa}. For each position, the inferred metallicity value results from the median of the metallicities determined by the respective authors for several hundred stars. We estimated an uncertainty of $\sim 0.05$ dex on the data points. This uncertainty should also account for any differences that may exist between the fields above and below the Galactic disc. The literature values at lowest latitudes, obtained by \citet{Zoccali:2008aa}, have a metallicity that is compatible within the uncertainties with the values obtained here. This supports a flat metallicity gradient in the inner parts of the bulge. Assuming that the Galactic bulge has probably formed from evolutionary processes of the metal-poor thick disc and the metal-rich thin disc, its metallicity gradient may result from the changing relative weights of these two components \citep[see][]{Babusiaux:2010aa,Garcia-Perez:2018aa}.  The different scales heights of $z^t_{thin\ disk} = 300\pm 50$ pc and $z^t_{thick\ disk} = 900\pm 180$ pc near the location of the Sun \citep{Bland-Hawthorn:2016aa} and the higher metallicity in the thin disc would produce the measured metallicity gradient.  Our fields are at low latitudes where the contribution from the thick disc will be practically constant, in agreement with a flat gradient in the inner few hundred parsecs.

\section{Summary}

We present deep, $0.2''$ angular resolution  $JHK_s$ photometry of two fields in the inner bulge, of size  7.95$'$  $\times$ 3.43$'$  at vertical separations of only $0.4^{\circ}$ and $0.6^{\circ}$ ($\sim$ 60 and $\sim$ 90 pc) from the Galactic centre.  Based on data from the  GALACTICNUCLEUS survey, we could thus overcome the high extinction and crowding in the inner bulge of the Milky Way and present deep studies of the CMDs and LFs of stars in the innermost bulge, where no comparable data from previous studies exist. We  identified the RGBB in the innermost bulge regions of the Milky Way. Thanks to its dependence on metallicity and age, we were able to constrain the properties of the stellar population using BaSTI isochrones. For the two fields, we obtained best fits for a single, old ($\sim12.8 \pm 0.6$\,Gyr) stellar population with a metallicity Z = $0.040 \pm 0.003$ dex. Given that the age of the stellar population was only measured indirectly in this work through a fit of model isochrones, there may be a significant bias. In particular, it is not easily conceivable how a large stellar population almost as old as the Universe could achieve super-solar metallicities. Nevertheless, given the results of our analysis, we can conservatively state that the population in the inner bulge, close to the nuclear stellar disc, is at least as old as or older than the age of the bulge population measured by other authors at larger latitudes \citep[e.g.,][]{Zoccali:2003aa,Clarkson:2008kx,Freeman:2008aa}. In this context, we would like to point out that our findings show that observations of the inner bulge of the Milky Way could be used to test and improve stellar evolutionary models for high metallicities.

We did not need to assume the contribution of any young stellar population ($\sim$ 5 Gyr) to obtain a satisfactory fit of the data, as in previous studies \citep{Clarkson:2011ys,Renzini:2018aa}. On the other hand, comparing the obtained metallicity in the studied fields with previous measurements at $b\approx2^{\circ}$, we obtained that our result favours a low-metallicity gradient that is compatible with a flattening of it in the inner regions.


 

As a secondary result, we found that the extinction index for the two bulge fields is consistent with the one  derived in \citet{Nogueras-Lara:2018aa},  $\alpha=2.30\pm0.08$, which indicates within the measurement uncertainties  that  the extinction curve across the Galactic centre region does not vary significantly. 

Finally, the stellar population of the regions in the inner bulge appears to be different from the population found at lower latitudes (in the NB), which is compatible with a continuous star formation history with recent stellar bursts \citep[e.g.][]{Figer:2004fk,Pfuhl:2011uq}.
Future investigations of a larger field at same spatial resolution will be helpful to map the metallicity and age distribution of the bulge population at low latitudes.

  \begin{acknowledgements}
      This work has made use of BaSTI web tools. The research leading to these results has received funding from
      the European Research Council under the European Union's Seventh
      Framework Programme (FP7/2007-2013) / ERC grant agreement
      n$^{\circ}$ [614922]. This work is based on observations made with ESO
      Telescopes at the La Silla Paranal Observatory under programmes
      IDs 195.B-0283 and 091.B-0418. We thank the staff of
      ESO for their great efforts and helpfulness. F. N.-L. acknowledges financial support from a MECD pre-doctoral contract, code FPU14/01700. F. N. acknowledges financial support through Spanish grants ESP2015-65597-C4-1-R and ESP2017-86582-C4-1-R (MINECO/FEDER).
\end{acknowledgements}

\bibliography{../../../BibGC}

\end{document}